\documentclass[aps,prl,twocolumn,showpacs,preprintnumbers,amsmath,amssymb,superscriptaddress]{revtex4}%

\usepackage{graphicx}%
\usepackage{dcolumn}
\usepackage{amsmath}
\usepackage{color}
\usepackage{bm}

\begin{document}


\title{Evidence for complex order parameter in La$_{1.83}$Sr$_{0.17}$CuO$_4$}

\author{R.~Khasanov}
 \affiliation{Physik-Institut der Universit\"{a}t Z\"{u}rich,
Winterthurerstrasse 190, CH-8057 Z\"urich, Switzerland}
\author{A.~Shengelaya}
 \affiliation{Physics Institute of Tbilisi State University,
Chavchadze 3, GE-0128 Tbilisi, Georgia } 
\author{A.~Maisuradze}
 \affiliation{Physik-Institut der Universit\"{a}t Z\"{u}rich,
Winterthurerstrasse 190, CH-8057 Z\"urich, Switzerland} 
\author{F.La~Mattina}
 \affiliation{Physik-Institut der Universit\"{a}t Z\"{u}rich,
Winterthurerstrasse 190, CH-8057 Z\"urich, Switzerland} 
%
%
\author{A.~Bussmann-Holder}
\affiliation{Max-Planck-Institut f\"ur Festk\"orperforschung,
Heisenbergstrasse 1, D-70569 Stuttgart, Germany}
\author{H.~Keller}
\affiliation{Physik-Institut der Universit\"{a}t Z\"{u}rich,
Winterthurerstrasse 190, CH-8057 Z\"urich, Switzerland}
\author{K.A.~M\"uller}
\affiliation{Physik-Institut der Universit\"{a}t Z\"{u}rich,
Winterthurerstrasse 190, CH-8057 Z\"urich, Switzerland} 

\begin{abstract}
The in-plane magnetic field penetration depth ($\lambda_{ab}$) in
single-crystal La$_{1.83}$Sr$_{0.17}$CuO$_4$ was investigated by
means of the muon-spin rotation ($\mu$SR) technique. The temperature
dependence of $\lambda^{-2}_{ab}$ has an inflection point around
10-15~K, suggesting the presence of two superconducting gaps: a
large gap ($\Delta_1^d$) with $d-$wave and a small gap
($\Delta_2^s$) with $s-$wave symmetry. The zero-temperature values
of the gaps at $\mu_0H=0.02$~T were found to be
$\Delta_1^d(0)=8.2(2)$~meV and $\Delta_2^s(0)=1.57(8)$~meV.
\end{abstract}
\pacs{76.75.+i, 74.72.Dn, 74.25.Ha}

\maketitle

It is mostly believed that the order parameter in cuprate
high-temperature superconductors (HTS) has purely $d-$wave symmetry,
as indicated by {\it e.g.} tricrystal experiments \cite{Tsuei94}.
There are, however, a wide variety of experimental data that support
$s$ or even more complicated  types of symmetries ($d+s$, $d+is$
{\it etc.}) \cite{Deutscher05}. In order to solve this controversy
M\"uller suggested the presence of two superconducting condensates
with different symmetries ($s-$ and $d-$wave) in HTS
\cite{Muller95,MullerKeller97}. This idea was generated partly
because two gaps were observed in $n-$type SrTiO$_3$
\cite{Binnig80}, the first oxide in which superconductivity was
detected. In addition, it is known that the two-order parameter
scenario leads to a substantial enhancement of the superconducting
transition temperature in comparison to a single-band model
\cite{Suhl59,Bussmann-Holder04}. The two-band model was successfully
used to explain superconductivity in MgB$_2$ \cite{MgB2TwoGap} and
is considered also to be relevant to understand superconductivity in
HTS \cite{Kresin92,Bussmann-Holder04}.

Important information on the symmetry of the order parameter can be
obtained from magnetic field penetration depth ($\lambda$)
measurements. In particular, $\lambda(T)$, which reflects the
quasiparticle density of states available for thermal excitations,
admits to probe the superconducting gap structure. Measurements of
the field dependence of $\lambda$ allow to study the anisotropy of
the superconducting energy gap \cite{Sonier00} and, in the case of
two-gap superconductors, to obtain details on the relative
contribution of each particular gap as a function of magnetic field
\cite{Serventi04}.
In this letter we report a study of the in-plane magnetic
penetration depth ($\lambda_{ab}$) in slightly overdoped
single-crystal La$_{1.83}$Sr$_{0.17}$CuO$_4$ by means of the
muon-spin-rotation ($\mu$SR) technique. At low magnetic fields
($\mu_0H\lesssim0.3$~T) $\lambda_{ab}^{-2}(T)$ exhibits an
inflection point at $T\simeq10-15$K. We interpret this feature as a
consequence of the presence of two superconducting gaps, analogous
to double-gap MgB$_2$ \cite{Carrington03}. It is suggested that the
large gap ($\Delta_1^d(0)=8.2(2)$~meV) has $d-$ and the small gap
($\Delta_2^s(0)=1.57(8)$~meV)  $s-$wave symmetry. With increasing
magnetic field the contribution of $\Delta_2^s$ decreases
substantially, in contrast to an almost constant contribution of
$\Delta^d_1$. Both the temperature and the field dependences of
$\lambda^{-2}_{ab}$ were found to be similar to what was observed in
double-gap MgB$_2$ \cite{Serventi04,Carrington03}.


The La$_{1.83}$Sr$_{0.17}$CuO$_4$ single-crystal was grown by the
travelling solvent floating zone technique \cite{Nakano98}. The
transition temperature $T_c$ and the width of the superconducting
transition at $\mu_0H\simeq0$~T were found to be 36.2~K and 1.5~K,
respectively \cite{Gilardi02}.
The $\mu$SR experiments were performed at the $\pi$M3 beam line at
the Paul Scherrer Institute (Villigen, Switzerland). The sample was
field cooled from above $T_c$ to 1.6~K in a series of fields ranging
from 20~mT to 0.64~T. The sample was aligned such that the $c$-axis
was parallel (within 1 degree, as measured by Laue x-ray
diffraction) to the external magnetic field.
In the transverse-field geometry the local magnetic field
distribution $P(B)$ probed by $\mu$SR inside the superconducting
sample in the mixed state is determined by  the coherence length
$\xi$ and the magnetic field penetration depth $\lambda$. In extreme
type II superconductors ($\lambda\gg\xi$) $P(B)$ is almost
independent of $\xi$, and the second moment of $P(B)$ is
proportional to $1/\lambda^4$ \cite{Brandt88}.
\begin{figure}[htb]
\includegraphics[width=1.0\linewidth]{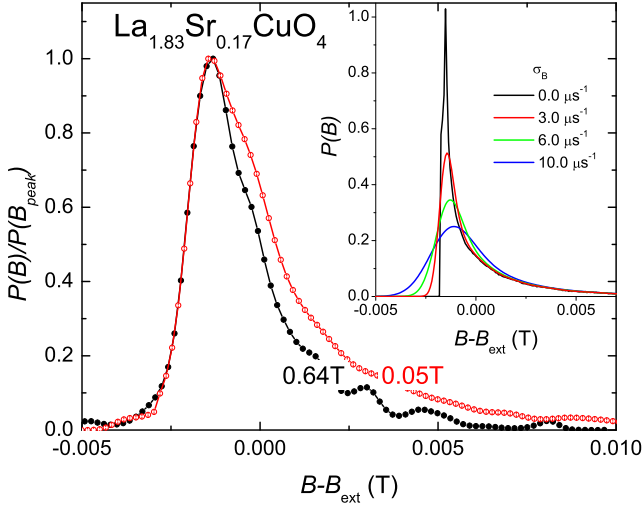}
%
\caption{Local magnetic field distribution $P(B)$ in the mixed state
of single-crystal La$_{1.83}$Sr$_{0.17}$CuO$_4$ ($T=1.7$~K,
field-cooled) normalized to their maximum value at $B=B_{peak}$ for
0.05~T and 0.64~T. The inset shows theoretical $P(B)$ distributions
($\lambda=220$~nm, $\xi=2$~nm, and $\mu_0H=0.05$~T) for different
values of the smearing parameter $\sigma_B=0$, 3, 6, and 10~$\mu
s^{-1}$.
 }
 \label{fig:absence_of_pinning}
\end{figure}

Figure~\ref{fig:absence_of_pinning} shows the magnetic field
distributions $P(B)$ for single-crystal
La$_{1.83}$Sr$_{0.17}$CuO$_4$ at $T=1.7$~K obtained by means of
the maximum entropy Fourier transform technique.
In order to extract the second moment of $P(B)$ we used a similar
procedure as described in Ref.~[\onlinecite{Khasanov05}]. All
$\mu$SR time spectra
were fitted by a three component expression:
\begin{equation}
P(t)= \sum_{i=1}^3A_i \exp(-\sigma_i^2t^2/2) \cos(\gamma_{\mu}B_i
t+\phi) \; .
\label{eq:gauss}
\end{equation}
Here $A_i$, $\sigma_i$, and $B_i$ are the asymmetry, the relaxation
rate, and the mean field of the ith component, and $\phi$ is the
initial phase of the muon-spin ensemble. The first and the second
moments of $P(B)$ are \cite{Khasanov05}:
\begin{equation}
\langle B \rangle=\sum_{i=1}^3{A_i B_i \over A_1+A_2+A_3} \;
\label{eq:B_mean}
\end{equation}
and
\begin{equation}
\langle \Delta B^2
\rangle=\frac{\sigma^2}{\gamma^2_\mu}=\sum_{i=1}^3{A_i \over
A_1+A_2+A_3} \left[ (\sigma_i/\gamma_{\mu})^2 +[B_i- \langle B
\rangle]^2 \right] ,
\label{eq:dB}
\end{equation}
where $\gamma_{\mu} = 2\pi\times135.5342$~MHz/T is the muon
gyromagnetic ratio. The superconducting part of the square root of
the second moment ($\sigma_{sc}\propto\lambda^{-2}_{ab}$) was then
obtained by subtracting the nuclear moment contribution
($\sigma_{nm}$) measured at $T>T_c$ according to
$\sigma_{sc}^2=\sigma^2 - \sigma_{nm}^2$ \cite{Khasanov05}.
To ensure that the increase of $\sigma$ below $T_c$ is attributed
entirely to the vortex lattice, zero-field $\mu$SR experiments were
performed. The experiments show no evidence for static magnetism in
La$_{1.83}$Sr$_{0.17}$CuO$_4$ down to 1.7~K.

\begin{figure}[htb]
\includegraphics[width=1.0\linewidth]{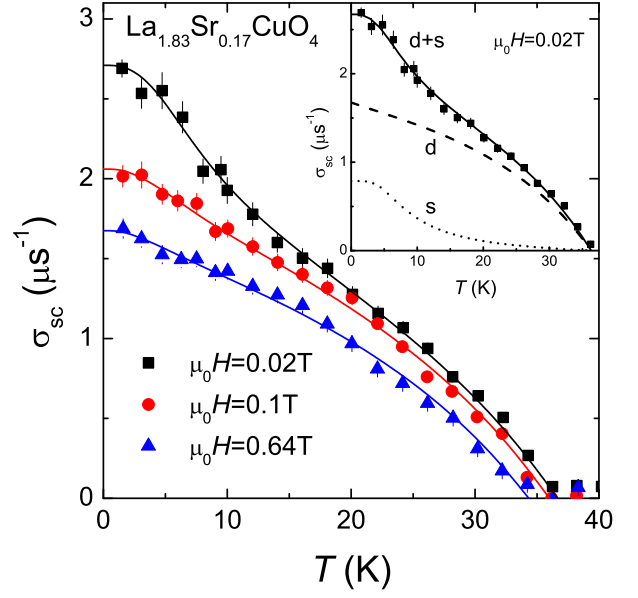}
%
\caption{ Temperature dependence of $\sigma_{sc}\propto
\lambda^{-2}_{ab}$ of single-crystal La$_{1.83}$Sr$_{0.17}$CuO$_4$
measured at 0.02~T, 0.1~T, and 0.64~T (field-cooled). Lines in the
main figure and in the inset represent the fit with the two gap
model [Eq.~(\ref{eq:sigma_two-gap})]. In the inset the contributions
from the large $d-$wave gap and the small $s-$wave gap entering
Eq.~(\ref{eq:sigma_two-gap}) are shown separately. See text for
details. }
 \label{fig:lambda_vs_T}
\end{figure}

In Fig.~\ref{fig:lambda_vs_T} we plot the temperature dependences of
$\sigma_{sc}\propto\lambda^{-2}_{ab}$  for $\mu_0 H=0.02$~T, 0.1~T,
and 0.64~T (for clarity, data for 0.05~T and 0.3~T are not shown).
Most importantly, around 10-15~K an inflection point appears. It is
well pronounced at $\mu_0H=0.02$~T and almost absent at
$\mu_0H=0.64$~T.
In Ref.~[\onlinecite{Xiang96}] it was pointed out that an inflection
point in $\lambda^{-2}(T)$ may appear in superconductors with two
weakly coupled superconducting bands. Indeed, in MgB$_2$, where the
$\sigma-$ and $\pi-$bands are almost decoupled, an upward curvature
of $\lambda^{-2}(T)$, similar to the one observed for
La$_{1.83}$Sr$_{0.17}$CuO$_4$ at $\mu_0H=0.02$~T
(Fig.~\ref{fig:lambda_vs_T}), was detected (see {\it e.g.}
\cite{Carrington03}). Thus, in analogy to MgB$_2$, we analyze our
data by assuming that $\sigma_{sc}$ is a linear combination of two
terms \cite{Niedermayer02,Kim02}:
\begin{equation}
\sigma_{sc}(T)/\sigma_{sc}(0)=\omega\cdot\delta\sigma(\Delta_1(0),T)+
(1-\omega)\cdot\delta\sigma(\Delta_2(0),T).
 \label{eq:sigma_two-gap}
\end{equation}
Here $\Delta_1(0)$ and $\Delta_2(0)$ are the zero-temperature values
of the large and the small gap, respectively, and $\omega$
($0\leq\omega\leq1$) is the weighting factor which measures their
relative contributions to $\lambda^{-2}$. Note, that in contrast to
MgB$_2$ where both gaps are isotropic, in HTS at least one gap has
$d-$wave symmetry \cite{Tsuei94}. Concerning the symmetry of the
second gap, however, the situation is unclear. Based on the
observation of a substantial $s-$wave contribution to the
superconducting order parameter by Andreev reflection experiments
\cite{Deutscher05} and on the analysis of tunnelling data
\cite{Muller95}, we assume that the second gap has isotropic
$s-$wave symmetry. Thus, for the contribution to $\sigma_{sc}$
arising from the $s-$wave gap we used the standard relation
\cite{Kim02}:
\begin{equation}
\delta\sigma(T,\Delta^s(0))=  1+
2\int_{\Delta^s(T)}^{\infty}\left(\frac{\partial f}{\partial
E}\right)\frac{E}{\sqrt{E^2-\Delta^s(T)^2}}\  dE.
 \label{eq:sigma-s}
\end{equation}
Here $f=(1+\exp(E/k_BT))^{-1}$ is  the Fermi function, $k_B$ is the
Boltzman constant, and $\Delta^s(T)=\Delta^s(0)
\tilde{\Delta^s}(T/T_c)$ represents the temperature dependence of
the $s-$wave gap with the tabulated gap values
$\tilde{\Delta^s}(T/T_c)$ from \cite{Muhlschlegel59}. For the
$d-$wave gap contribution we take
$\Delta^d(T,\varphi)=\Delta^s(T)\cos(2\varphi)$ \cite{Deutscher05}
and
\begin{equation}
\delta\sigma(T,\Delta^d(0))=  1+
2\int_{0}^{2\pi}\int_{\Delta^d(T,\varphi)}^{\infty}\left(\frac{\partial
f}{\partial E}\right)\frac{E}{\sqrt{E^2-\Delta^d(T,\varphi)^2}}\
dEd\varphi.
 \label{eq:sigma-d}
\end{equation}

In order to determine the symmetry of the two gaps, the field-cooled
0.05~T data were analyzed within ''$d+s$`` and ''$s+d$`` scenarios
using Eq.~(\ref{eq:sigma_two-gap}). The analysis reveals for
''$d+s$``: $\Delta_1^d(0)=9.0(2)$~meV, $\Delta_2^s(0)=1.7(1)$~meV,
$\omega=0.69(3)$, and for ''$s+d$``: $\Delta_1^s(0)=6.2(2)$~meV,
$\Delta_2^d(0)=2.0(2)$~meV, $\omega=0.73(2)$. Comparison with
$\Delta(0)\simeq$10~meV obtained on a similar sample by tunnelling
experiments \cite{Oda00}, suggests that the large gap has $d-$wave
symmetry. Another argument in favor of a ''large`` $d-$wave gap
comes from the observation of a square vortex lattice in the same
crystal as used in this work in fields higher than 0.4~T
\cite{Gilardi02,Drew05}, where as shown below the contribution from
the large gap to $\sigma_{sc}$ is dominant. A square vortex lattice
is typical for $d-$wave superconductors \cite{Gilardi02}.

\begin{figure}[htb]
\includegraphics[width=0.9\linewidth]{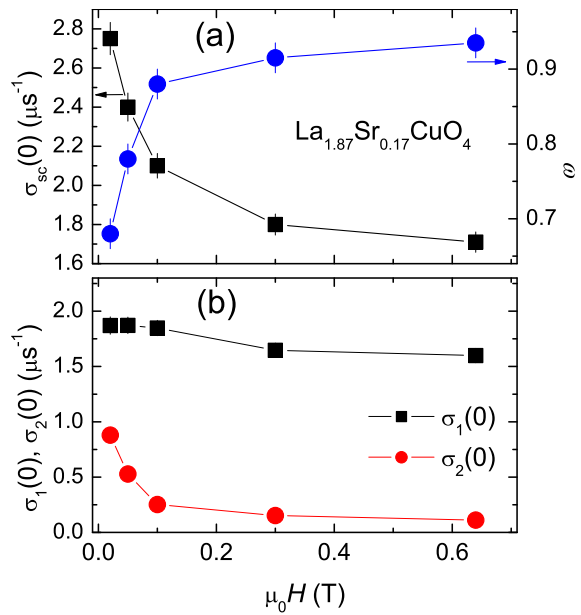}
%
\caption{ (a) -- Field dependences of $\sigma_{sc}(0)$ and $\omega$
for single-crystal  La$_{1.83}$Sr$_{0.17}$CuO$_4$ obtained from the
fit of Eq.~(\ref{eq:sigma_two-gap}) to the data (see
Table~\ref{Table:two-gap}). (b) -- Contribution from the large
[$\sigma_1(0)$] and the small [$\sigma_2(0)$] superconducting gap to
the total $\sigma_{sc}(0)$.
 }
 \label{fig:sigma_omega}
\end{figure}

The solid lines in Fig.~\ref{fig:lambda_vs_T} represent the global
fit of Eq.~(\ref{eq:sigma_two-gap}) to the data with contributions
from the large and the small gaps described by
Eqs.~(\ref{eq:sigma-d}) and (\ref{eq:sigma-s}), respectively. In the
analysis all the $\sigma_{sc}(T)$ curves (0.02, 0.05, 0.1, 0.3,
0.64~T) were fitted simultaneously with $\sigma_{sc}(0)$, $T_c$, and
$\omega$ as individual parameters for each particular data set.
$\Delta_1^d(0)$ and $\Delta_2^s(0)$ were assumed to scale linearly
with $T_c$ according to the relation $2\Delta(0)/k_BT_c=$~const. The
results are summarized
in Table~\ref{Table:two-gap} and Fig.~\ref{fig:sigma_omega}. %
\begin{table}[htb]
\caption[~]{\label{Table:two-gap} Summary of the two-gap analysis
for single-crystal  La$_{1.83}$Sr$_{0.17}$CuO$_4$. The meaning of
the parameters is -- $\mu_0H$: external magnetic field, $T_c$:
superconducting transition temperature, $\sigma_{sc}(0)$:
zero-temperature $\mu$SR relaxation rate, $\omega$: relative
weighting factor, $\Delta^d_1(0)$: $d-$wave gap, $\Delta_2^s(0)$:
$s-$wave gap. }
\begin{center}
\begin{tabular}{lcccccccc}\\ \hline
\hline
$\mu_0$H &$T_c$&$\sigma_{sc}(0)$&$\omega$&$\Delta_1^d(0)$&$\Delta_2^s(0)$\\
(T) &(K)&($\mu$s$^{-1}$)&&(meV)&(meV)\\
\hline
0.02&36.3(1)&2.71(8)&0.68(3)&8.2(1)&1.57(8)\\
0.05&36.1(1)&2.20(7)&0.78(2)&8.2(1)&1.56(8)\\
0.1&35.5(1)&2.07(7)&0.88(2)&8.0(1)&1.54(8)\\
0.3&34.7(1)&1.82(6)&0.92(2)&7.8(1)&1.50(7)\\
0.64&34.0(1)&1.71(5)&0.94(2)&7.7(1)&1.47(7)\\

 \hline \hline \\

\end{tabular}
   \end{center}
\end{table}
It is seen in Fig.~\ref{fig:sigma_omega}(a) that the decrease of
$\sigma_{sc}(0)$ is associated with an increase of the contribution
of the large gap to $\lambda^{-2}$. Similar field dependences of
$\omega$ and $\sigma_{sc}$ were observed in MgB$_2$ by small angle
neutron scattering \cite{Cubitt03}, point-contact spectroscopy
\cite{Gonelli02}, and $\mu$SR \cite{Serventi04} experiments. This
was explained by the fact that superconductivity within the weaker
$\pi$-band is suppressed at much lower fields than that within the
stronger $\sigma$-band \cite{Gonelli02}. As shown in
Fig.~\ref{fig:sigma_omega}(b) this is also the case for
La$_{1.83}$Sr$_{0.17}$CuO$_4$. Indeed, while the contribution from
the large gap [$\sigma_1(0)=\omega\cdot\sigma_{sc}(0)$] changes only
slightly, the contribution from the small gap
[$\sigma_2(0)=(1-\omega)\cdot\sigma_{sc}(0)$] decreases by almost an
order of magnitude in the field range $0<\mu_0H\leq0.64$~T
[Fig.\ref{fig:sigma_omega}(b)]. Thus, the temperature and the field
dependences of $\lambda^{-2}_{ab}$ in La$_{1.83}$Sr$_{0.17}$CuO$_4$
are similar to MgB$_2$, and consequently demonstrate the existence
of two gaps. This is the most obvious scenario, even though other
gap dependences cannot be fully ruled out.

It is important to emphasize that the observation of an inflection
point in $\lambda^{-2}(T)$ is not restricted to MgB$_2$ and the
particular sample used in this work. Indication of an inflection
point in $\lambda^{-2}(T)$ was also observed in hole-doped
YBa$_2$Cu$_3$O$_{7-\delta}$ \cite{Sonier00,Harshman04},
YBa$_2$Cu$_4$O$_8$ \cite{Panagopoulos98}, and
La$_{1.85}$Sr$_{0.15}$CuO$_4$ \cite{Luke97}, as well as in
electron-doped Pr$_{1.855}$Ce$_{0.145}$CuO$_{4–y}$ \cite{Skinta02}.
In Ref.~[\onlinecite{Harshman04}] the increase of the second moment
of $P(B)$ observed in YBa$_2$Cu$_3$O$_{7-\delta}$ at low
temperatures was attributed to pinning effects. In order to
investigate the role of pinning in our sample we compare the $P(B)$
distributions for 0.05~T and 0.64~T
(Fig.~\ref{fig:absence_of_pinning}) with theoretical $P(B)$ curves.
A standard way to account for pinning is to convolute the
theoretical $P(B)$ for an ideal vortex lattice (black line in the
inset of Fig.~\ref{fig:absence_of_pinning}) with a Gaussian
distribution of fields \cite{Brandt88a}:
\begin{equation}
P(B)=\frac{1}{\sqrt{2\pi}\sigma_B}\int\exp\left[-\frac{1}{2}\left(
\frac{B-B'}{\sigma_B}\right)^2\right]P_{id}(B')dB',
 \label{eq:pinning}
\end{equation}
where $\sigma_B$ is the width of the Gaussian distribution and
$P_{id}(B)$ is the field distribution for an ideal vortex lattice
\cite{Sonier00}. It was shown \cite{Brandt88a} that for a stiff
vortex lattice this convolution reflects how random disorder and
distortions due to flux line pinning influence the ideal
$P_{id}(B)$. The theoretical $P(B)$ profiles for $\sigma_B$=0, 3, 6,
and 10 $\mu s^{-1}$ are shown in the inset of
Fig.~\ref{fig:absence_of_pinning}. The direct comparison of the
$P(B)$ data for $\mu_0H$=0.05~T and 0.64~T with theoretical $P(B)$
profiles clearly demonstrates that pinning is not the main source of
the observed increase of the second moment of $P(B)$ at low
temperatures. Indeed, pinning leads to an almost symmetric (around
$B_{peak}$) broadening of $P(B)$ (see inset of
Fig.~\ref{fig:absence_of_pinning}), while the experimental $P(B)$
profiles very well coincide at low fields ($B<B_{peak}$). Deviations
only occur in the high-field tail of $P(B)$ ($B>B_{peak}$).

The obvious question which arises is where to locate the second
superconducting gap in La$_{2-x}$Sr$_x$CuO$_4$? The phase diagram of
cuprates is usually interpreted in terms of holes doped into the
planar Cu$d_{x^2-y^2}$-O$p_\alpha$ ($\alpha=x,y$) antibonding band.
In La$_{2-x}$Sr$_x$CuO$_4$ it is assumed that one hole per Sr atom
enters this band. However, recent $ab-initio$ calculations yielded
additional features appearing on doping of La$_{2-x}$Sr$_x$CuO$_4$
\cite{Perry02}. According to these calculations part of the holes
occupy the Cu$d_{3z^2-r^2}$-O$p_z$ orbitals. These results are
further supported by neutron diffraction data \cite{Bozin03},
showing that the doped holes indeed appear in both the planar and
the out-of-plane bands. In contrast to this finding, in
angle-resolved photoemission (ARPES) experiments on HTS only the
planar band was observed, suggesting a quasi-two-dimensional
electronic structure with negligible {\it intercell} coupling of
CuO$_2$-layers (see {\it e.g.} \cite{Damascelli03}). This is,
however, inconsistent with in-plane and out-of-plane penetration
depth measurements \cite{Xiang98}, optical conductivity
\cite{Tamasaku94}, and anisotropy parameter studies \cite{Hofer00}.
All these experiments demonstrate that with increasing  doping
cuprates become more and more three-dimensional. Recently a 3D Fermi
surface was observed in overdoped TlBa$_2$CuO$_{6+\delta}$
\cite{Hussey03}. In addition, a careful analysis of ARPES data
reveals that the finite dispersion of the energy bands along the
$z$-direction of the Brillouin zone ($k_z$ dispersion) naturally
induces an irreducible linewidth of the ARPES peaks which is
unrelated to any scattering mechanism \cite{Bansil05Sahrakorpi05}.
This implies that a single 2D band model is insufficient and
out-of-plane hybridized bands have to be incorporated.

In conclusion, we performed systematic $\mu$SR studies of the
in-plane magnetic penetration depth $\lambda_{ab}$ in single-crystal
La$_{1.83}$Sr$_{0.17}$CuO$_4$. Both, the magnetic field and the
temperature dependences of $\lambda^{-2}_{ab}$ were found to be
consistent with the presence of two gaps, analogous to MgB$_2$.
Accordingly, the experimental data were analyzed by assuming that
the large gap ($\Delta^d_1$) has $d-$wave and the small gap
($\Delta^s_2$) $s-$wave symmetry. The zero temperature values of the
superconducting gaps at $\mu_0H=0.02$~T were determined to be
$\Delta^d_1$(0)=8.2(2)~meV and $\Delta^s_2$(0)=1.57(8)~meV,
corresponding to $2\Delta^d_1(0)/k_BT_c=5.24(7)$ and
$2\Delta^s_2(0)/k_BT_c=1.00(5)$, respectively. The contribution of
the small gap to the superfluid density $\lambda^{-2}_{ab}$ was
found to decrease from 32(3)\% at $\mu_0H=0.02$~T to 6(2)\% at
$\mu_0H=0.64$~T. Further $\mu$SR investigation of the penetration
depth in La$_{2-x}$Sr$_x$CuO$_4$ at various doping levels are in
progress.

This work was partly performed at the Swiss Muon Source (S$\mu$S),
Paul Scherrer Institute (PSI, Switzerland). The authors are grateful
to N.~Momono, M.~Oda, M.~Ido and J.~Mesot for providing us the
La$_{1.83}$Sr$_{0.17}$CuO$_4$ single crystal, J.~Mesot for helpful
discussions, and A.~Amato, D.~Herlach and C.J.~Juul for assistance
during the $\mu$SR measurements. This work was supported by the
Swiss National Science Foundation, in part by the NCCR program
MaNEP, the EU Project CoMePhS, and the K.~Alex~M\"uller Foundation.

\end{document}